\documentclass[10pt,twocolumn]{article}

\usepackage{amsmath}
\usepackage{graphicx}
\usepackage{lineno}

\title{A Very Simple Approach for 3-D to 2-D Mapping}

\author{Sandipan Dey,
\\Anshin Software Pvt. Ltd.,
\\email: sandipand@anshinsoft.com
\and
Ajith Abraham,
\\IITA Professorship Program, School of Computer Science,
\\Yonsei University,
\\134 Shinchon-dong, Sudaemoon-ku, Seoul 120-749, Republic of Korea
\\email: ajith.abraham@ieee.org \and
Sugata Sanyal,
\\School of Technology \& Computer Science
\\Tata Institute of Fundamental Research
\\Homi Bhabha Road, Mumbai - 400005, INDIA
\\email: sanyal@tifr.res.in\\}

\usepackage{graphicx}

\begin{document}

\maketitle \thispagestyle{empty}

\section*{Abstract}

\noindent Many times we need to plot 3-D functions e.g., in
many scientific experiments. To plot this 3-D functions
on 2-D screen it requires some kind of mapping.
Though OpenGL, DirectX etc 3-D rendering
libraries have made this job very simple, still these
libraries come with many complex pre-operations
that are simply not intended, also to integrate these
libraries with any kind of system is often a tough
trial. This article presents a very simple method of
mapping from $3-D$ to $2-D$, that is free from any complex
pre-operation, also it will work with any graphics
systemwhere we have some primitive 2-D graphics
function. Also we discuss the inverse transform
and how to do basic computer graphics transformations
using our coordinate mapping system.

\section{Introduction}
\noindent We have a function $f : \Re^2 \rightarrow \Re$, and our intention
is to draw the function in $2-D$ plane. The function
$z = f (x, y)$ is a $2$-variable function and each tuple
$(x, y, f(x,y)) \in \Re^3$. Let's say we want to graphically
plot $f$ onto computer screen using a primitive graphics library (like Turbo C graphics), which
supports only the basic putPixel (to draw a pixel in $2-D$ screen)-like $2-D$ rendering function, but no
$3-D$ rendering; i.e., our graphics library's putPixel's domain is $\Re^2$ and it's not $\Re^3$.

\section{Proposed Approach}

\noindent We have a pictorial representation (figure $1$) of our $3-D$
to $2-D$ mapping system:

\begin{figure}[htbp]
    \centering
        \includegraphics[width = 8cm, height = 2.2cm]{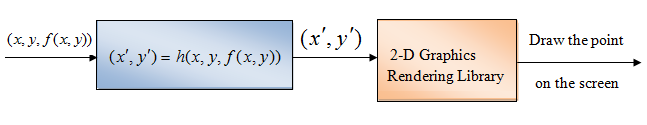}
        \caption{Basic Model of a simple $3-D$ to $2-D$ mapping system}
\end{figure} 

\noindent But, how the function $f$ should look like after the mapping and the plotting? Here we simulate 
the $3^{rd}$ coordinate (namely $Z$) in our $2-D$ x-y plane. We perform the $logical$ to $physical$ coordinate 
transform and everything by the map function $h$, which will basically turn out to be a $3\times2$ matrix. 
The basic mapping technique is shown in fig. $2$, which we are going to explain shortly. \\

\noindent We start with our Origin $O$ (i.e. $(0,0,0)$) mapped to $(x_0, y_0)$ screen coordinate, so that
we have the following equations,

\begin{eqnarray}
	x\prime = x_0 + y - x.sin(\theta) \nonumber \\
	y\prime = y_0 - z + x.cos(\theta)
	\label{eq1:}
\end{eqnarray}

i.e., we have our $3-D$ to $2-D$ transformation matrix:
\begin{eqnarray}
M_{3\times 2}=\bordermatrix{
&&\cr   
& -sin(\theta) & cos(\theta) \cr   
& 1 & 0 \cr   
& 0 & -1 \cr   
}
\end{eqnarray}

\begin{figure}[htbp]
    \centering
        \includegraphics[width = 8cm, height = 7.5cm]{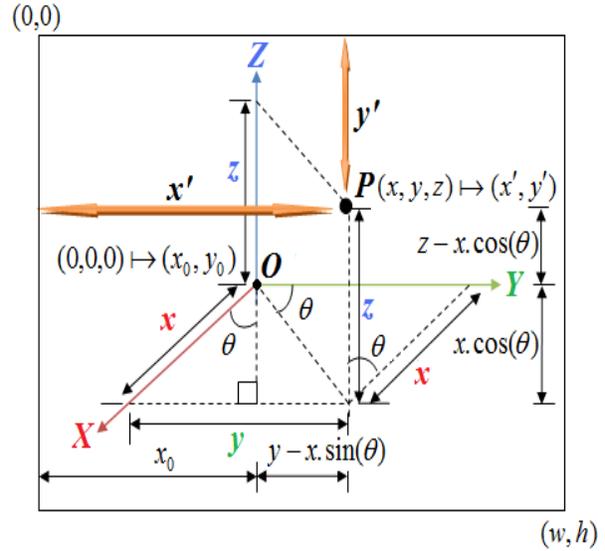}
        \caption{The coordinate mapping $h$: logical coordinate $(x,y,z) \mapsto$ physical coordinate $(x\prime, y\prime)$, given the 
        logical origin O $\mapsto$ physical origin $(x_0, y_0)$}
\end{figure} 

\noindent Again we have change of origin (shifting) by the matrix (vector) $O_{2D} = [x_0\; y_0]$ so that $O_{2D} +P_{3D} \times M_{3\times 2} =
P_{2D}$, here $\times$ denotes matrix multiplication and $+$ denotes matrix addition, the $3$-tuple $P_{3D} = [x\;y\;z]$, the
$2$-tuple $P_{2D} = [x\prime\; y\prime]$, i.e.,
\begin{eqnarray}
[x_0\; y_0] + [x\; y\; z].\bordermatrix{
&&\cr   
& -sin(\theta) & cos(\theta) \cr   
& 1 & 0 \cr   
& 0 & -1 \cr   
}=[x\prime\; y\prime]
\end{eqnarray}

\noindent By default we keep the angle between $X-$axis and $Z-$axis $=\theta=\frac{\pi}{4}$, that one can
change if required, but with the following inequality strictly satisfied: $0 < \theta < \frac{\pi}{2}$. \\

\noindent One can optionally use a compression factor to control the dimension along $Z-$axis by a compression
factor $\rho_z$ and slightly modifying the equations

\begin{eqnarray}
	x\prime = x_0 + y - x.sin(\theta) \nonumber \\
	y\prime = y_0 - \rho_z.z + x.cos(\theta)
	\label{eq1:}
\end{eqnarray}
 
\noindent Obviously, $0.0 <\rho_z < 1.0$. By default we take $\rho_z = 1.0$.

\section{Sample output surfaces drawn using the above mapping}
\noindent Following surfaces (Fig. 3 and Fig. 4) are drawn in Turbo C++ version 3.0 (BGI graphics) using the above simple $3-D$ to $2-D$ mapping.

\noindent Also we used the same technique for a 3-D plot in java applet, the outputs are shown in Fig. 5. We used double-buffering technique for smooth drawing (copying in an alternate buffer and copying it back to original graphic device context only when plot finishes to avoid flickering) for better performance (since drawing pixel by pixel is a bit time consuming). Also, we used color code mapping to have gradient effect in our plot. 

\begin{figure}[htbp]
    \centering
        \includegraphics[width = 8cm, height = 15cm]{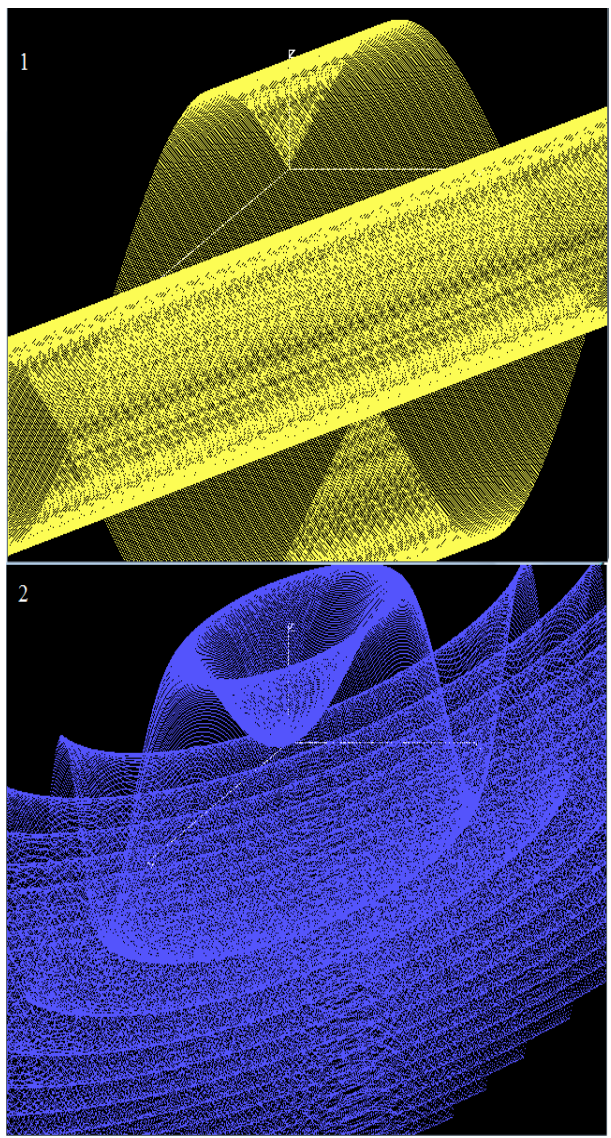}
        \caption{1) $sin(x+y)$ 2) $sin(x^2+y^2)$ functions drawn in TurboC++ Version 3.0 (BGI Graphics) using the $3-D$ to $2-D$ mapping, 
        with proper scaling}
\end{figure} 

\begin{figure}[htbp]
    \centering
        \includegraphics[width = 8cm, height = 15cm]{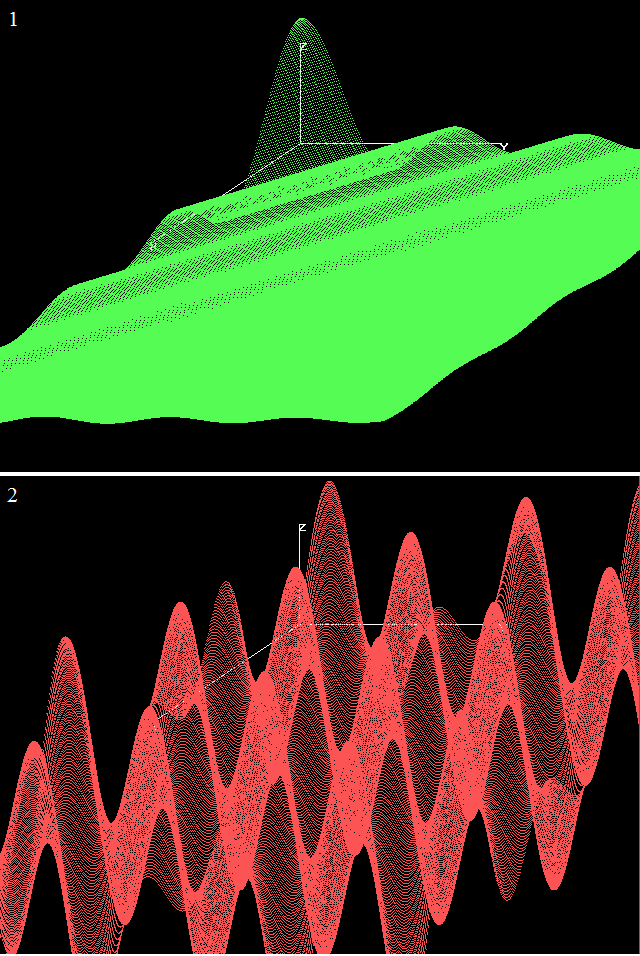}
        \caption{1) $\frac{sin(x+y)}{x+y}$ 2)$sin(\frac{x-y}{a})cos(\frac{x+y}{b})$ functions drawn in TurboC++ Version 3.0 (BGI 
        Graphics) using the $3-D$ to $2-D$	mapping, with proper scaling}
\end{figure} 

\begin{figure}[htbp]
    \centering
        \includegraphics[width = 8cm, height = 15cm]{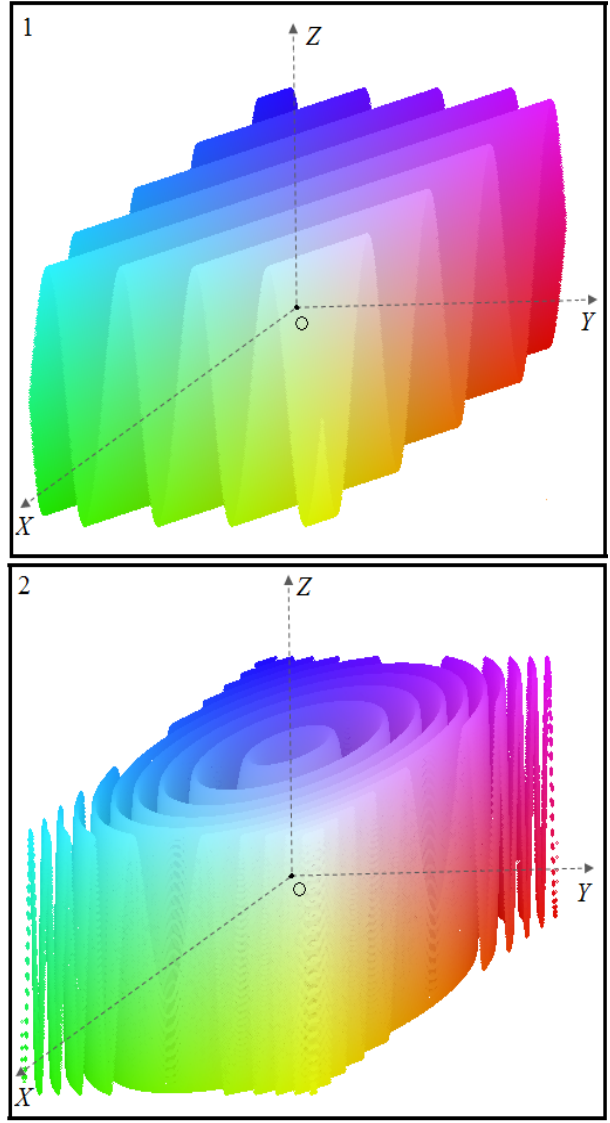}
        \caption{1) $sin(x+y)$ 2) $sin(x^2+y^2)$ functions drawn in java applet $2D$ graphics using the $3-D$ to $2-D$ mapping, 
        with proper scaling and color}
\end{figure} 

\section{Inverse Transformation - Obtaining the original $3-D$ coordinates from the transformed $2-D$ coordinates}

\noindent Here, our transformation function (matrix) is defined by Eqn. $(1)$.
As we can see, it is impossible to (uniquely) re-convert and obtain the original set of coordinates, namely $(x, y, z)$,
because we have $3$ unknowns and $2$ equations. So, in order to be able to get the original coordinates back, 
we at least need to store $3$ tuples as result of the transformation, for instance, $(x, y, z) \mapsto (x\prime, y\prime, z)$,
the $z$-coordinate being stored only to get the inverse transform $(x\prime, y\prime, z) \mapsto (x, y, z)$ and the $(x\prime, y\prime)$ pair is used to plot the point. Hence, in order to get the inverse transformation, we need to solve the equations for
$x$, $y$, since we already know $z$, we have $2$-equations and $2$ unknown variables: 

\begin{eqnarray}
	y - x.sin(\theta) = x\prime - x_0  \nonumber \\
	x.cos(\theta) = y\prime - y_0 + z 
	\label{eq3:}
\end{eqnarray}

\noindent solving the above $2$ equations we get,

\begin{eqnarray}
	x = (y\prime - y_0 + z ).sec(\theta) \nonumber \\
	y = x\prime - x_0 + (y\prime - y_0 + z ).tan(\theta) 
	\label{eq3:}
\end{eqnarray}

\noindent Put it in another way, our transformation matrix
is a $3\times 2$ matrix and is performed by Eqn. $(2)$ since a
non-square matrix, no question of existence of its inverse. 
So, in order to be able to get the inverse transform as well, we need a $3\times 3$ invertible square
matrix, e.g.,
\begin{eqnarray}
M_{3\times 3}=\bordermatrix{
&&&\cr   
& -sin(\theta) & cos(\theta) & 0 \cr   
& 1 & 0 & 0 \cr   
& 0 & -1 & 1 \cr   
}
\end{eqnarray}

\noindent with 

\begin{eqnarray}
Det(M_{3\times 3})=det\bordermatrix{
&&&\cr   
& -sin(\theta) & cos(\theta) & 0 \cr   
& 1 & 0 & 0 \cr   
& 0 & -1 & 1 \cr   
} \nonumber \\
=\bordermatrix{
&&\cr   
& -sin(\theta) & cos(\theta) \cr   
& 1 & 0 \cr   
} 
= -cos(\theta)
\end{eqnarray}

\noindent Now, $0 < \theta < \frac{\pi}{2}$, hence $cos(\theta) \neq 0$, hence $Det(M_{3\times3}) \neq 0$, i.e. the matrix is 
non-singular and the inverse exists. 

\begin{eqnarray}
[x_0\; y_0\; 0] + [x\; y\; z]\bordermatrix{
&&&\cr   
& -sin(\theta) & cos(\theta) & 0 \cr   
& 1 & 0 & 0 \cr   
& 0 & -1 & 1 \cr   
}=[x\prime\; y\prime\; z]
\end{eqnarray}

\noindent But, we have,
\begin{eqnarray}
Inv(M_{3\times 3}) = M_{3 \times 3}^{-1} = \frac{Adj(M_{3 \times 3})}{Det(M_{3 \times 3})} \\
Det(M_{3 \times 3}) \neq 0 \nonumber
\end{eqnarray}

\noindent and,

\begin{eqnarray}
Adj(M_{3\times 3})=\bordermatrix{
&&&\cr   
& 0 & -cos(\theta) & 0 \cr   
& -1 & -sin(\theta) & 0 \cr   
& -1 & -sin(\theta) & -cos(\theta) \cr   
}
\end{eqnarray}

\noindent Hence,

\begin{eqnarray}
Inv(M_{3\times 3})=\bordermatrix{
&&&\cr   
& 0 & 1 & 0 \cr   
& sec(\theta) & tan(\theta) & 0 \cr   
& sec(\theta) & tan(\theta) & 1 \cr   
}
\end{eqnarray}

\noindent here, $cos(\theta) \neq 0$. \\

\noindent Hence, the inverse transform is:

\begin{eqnarray}
[x\; y\; z].\bordermatrix{
&&&\cr   
& -sin(\theta) & cos(\theta) & 0 \cr   
& 1 & 0 & 0 \cr   
& 0 & -1 & 1 \cr   
} 
=[x\prime\; y\prime\; z] - [x_0\; y_0\; 0] \\
\Rightarrow [x\; y\; z] = [x\prime - x_0\; y\prime - y_0\; z]
\bordermatrix{
&&&\cr   
& 0 & 1 & 0 \cr   
& sec(\theta) & tan(\theta) & 0 \cr   
& sec(\theta) & tan(\theta) & 1 \cr   
} 
\end{eqnarray}

\begin{eqnarray}
[x\; y\; z] = [(y\prime - y_0 + z)sec(\theta)\; x\prime - x_0 + (y\prime - y_0 + z)tan(\theta)\; z]
\end{eqnarray}

\noindent This exactly matches with our previous derivation.

\section{Rotation and Affine Transformations}

\noindent A point in $3-D$, after being mapped to $2-D$ screen 
following the above mapping procedure, may be
required to be transformed using standard computer
graphics transformations (translation, rotation
about an axis etc). But in order to undergo such a
graphics transformation and to show the point back
to the screen after the transformation, it needs to
go through the following steps in our previously-described
coordinate mapping system:

\begin{enumerate}
	\item First obtain the inverse coordinate transformation
	to obtain the original $3-D$ coordinates from
	the mapped $2-D$ coordinates.
	\item Multiply the $3-D$ coordinate matrix by proper
	graphics transformation matrix in order to
	achieve graphical transformation.
	\item Use the same $3-D$ to $2-D$ map again to plot the
	point onto the screen.
\end{enumerate}

\noindent These steps can be mathematically represented as:
\begin{enumerate}
	\item $P_{3D}=P_{2D}\times (M_{3 \times 3})^{-1}$
	\item $P\prime_{3D}=P_{3D}\times (T_{3 \times 3})$
	\item $P\prime_{2D}=P\prime_{3D}\times (M_{3 \times 3})$
\end{enumerate}

\noindent Or, by a single line expression,
\begin{eqnarray*}
	P\prime_{2D}=\left(\left(\left(P_{2D}\times (M_{3 \times 3})^{-1}\right) \times (T_{3 \times 3})\right) \times (M_{3 \times 3})\right)
	\label{eq4:}
\end{eqnarray*}

\noindent Here, as before $\times$ denotes matrix multiplication,
where $T_{3\times 3}$ denotes the traditional graphics transformation matrix. \\

\noindent But, since we know the fact that matrix multiplication
is associative, we have,

\begin{eqnarray}
	P\prime_{2D}=\left(\left(\left(P_{2D}\times (M_{3 \times 3})^{-1}\right) \times (T_{3 \times 3})\right) \times (M_{3 \times 3})\right) 
	\nonumber \\
	= P_{2D}\times M_{3 \times 3}^{-1} \times T_{3 \times 3} \times M_{3 \times 3} \nonumber \\
	\Rightarrow P\prime_{2D} = P_{2D}\times M\prime, \nonumber \\
	\mbox{ where } M\prime = M_{3 \times 3}^{-1} \times T_{3 \times 3} 
	\times M_{3 \times 3}	
	\label{eq4:}
\end{eqnarray}

\noindent Hence, using this simple technique, we can escape the $3$ successive matrix multiplications every-time a point on screen needs to transformed: instead we can pre-compute the matrix $M\prime = M_{3 \times 3}^{-1} \times T_{3 \times 3} \times M_{3 \times 3}$. \\

\noindent This matrix $M\prime$ is needed to be computed once for a given graphics transformation (e.g., rotation about
an axis) and applied to all points on the screen, so that using a single matrix multiplication thereafter
any point on the screen can undergo graphics transformation, by, $P\prime_{2D} = P_{2D}\times M\prime$, where $P_{2D}$ represents
the point mapped before transformation $T_{3\times 3}$ and $P\prime_{2D}$ is the point re-mapped after the transformation,
as obvious. \\

\noindent Hence, using the above tricks we are able to make the transformation more computationally efficient.
Moreover, if a transformation is needed to be applied simultaneously, we can use the property
$M_{3 \times 3}^{-1} \times (T_{3 \times 3})^n \times M_{3 \times 3}=(M_{3 \times 3}^{-1} \times T_{3 \times 3} \times M_{3 \times 3})^n$,
where $(T_{3\times 3})^n$ denotes (n times, n is a positive integer) simultaneous matrix multiplication of $T_{3\times3}$.
Letfs say we have already undergonea$T_{3\times3}$ transformation, so that we have already computed $M\prime = M_{3 \times 3}^{-1} \times T_{3 \times 3} \times M_{3 \times 3}$, and let's say we also have frequent simultaneous $T_{3\times3}^n$ transformation. In
order to undergo a $T_{3\times3}^n$ transformation, we first need to compute the matrix $T_{3\times3}^n$, then we need
to compute our new matrix $M\prime\prime = M_{3 \times 3}^{-1} \times(T_{3 \times 3})^n \times M_{3 \times 3}$, so we need total $n+2$ matrix multiplications, every-time we want a $T_{3\times3}^n$ transform, for each $n$. \\

\noindent But if we have computed $M_{3 \times 3}^{-1} \times (T_{3 \times 3})^n \times M_{3 \times 3}$ initially,
here the trick is that we can reuse this it to compute our new matrix in the following manner: \\ \\
$M\prime\prime=M_{3 \times 3}^{-1} \times (T_{3 \times 3})^n \times M_{3 \times 3}=(M_{3 \times 3}^{-1} \times T_{3 \times 3} \times M_{3 \times 3})^n=(M\prime)^n$. \\

\noindent Here we need not compute $T_{3\times3}^n$ and $M\prime\prime$ everytime,
instead we need to compute $(M\prime)^n$ only (that can be incremental multiplication to increase efficiency).

\section{Conclusion}
\label{conclu}
\noindent This article presented a very simple method of mapping
from $3-D$ to $2-D$, that is free from any complex
pre-operation. The proposed technique works with
any graphics system where we have some primitive
$2-D$ graphics function. We also discussed the inverse
transform and how to do basic computer graphics
transformations using our coordinate mapping system.

\section{References}
\begin{enumerate}
\item David F. Rogers, J. Alan Adams, Mathematical
Elements for Computer Graphics, McGraw-
Hill
\item  David F. Rogers, Procedural elements for computer
Graphics, United States Naval Academy,
Annapolis, MD
\item  Dave Shreiner, Mason Woo, Jackie Neider,
Tom Davis, OpenGL Programming Guide, The
Official Guide to Learning OpenGL, Version
1.4, Fourth Edition.
\item  Ken Turkowski, The Use of Coordinate Frames
in Computer Graphics, Graphics Gems I, Academic
Press, 1990, pp. 522-532.
\item  Ken Turkowski, Fixed-Point Trigonometry with
CORDIC Iterations, Graphics Gems I, Academic
Press, 1990, pp. 494-497.
\item  C.M. Ng, D.W. Bustard, A New Real Time Geometric
TransformationMatrix and its Efficient
VLSI Implementation, Computer Graphics Forum,
Volume 13 Page 285
\end{enumerate}

\end{document}